\documentstyle[12pt]{article}
\begin{document}\thispagestyle{empty}
\begin{center} \LARGE \tt \bf {Stringent limits for the magnetic field from early universe dynamos in $RF^{2}$ cosmology with torsion}
\end{center}

\vspace{1.0cm}

\begin{center}
{\large By L.C. Garcia de Andrade\footnote{Departamento de F\'{\i}sica Te\'{o}rica - IF - UERJ - Rua S\~{a}o Francisco Xavier 524, Rio de Janeiro, RJ, Maracan\~{a}, CEP:20550.e-mail:garcia@dft.if.uerj.br}}
\end{center}

\begin{abstract} Earlier Bamba et al [JCAP (2012)] have obtained cosmological magnetic fields in teleparallel torsion theories of gravity that are not compatible with galactic dynamos. This result agrees with previous ones obtained by the author which shows [Phys Lett B (2012)] that anti-dynamo generalised theorem to torsion theories forbides such kind of dynamos to explain galactic magnetic fields of the order of ${\mu}$G. More recently the author has suggested [IJAA (2012)] that a sort of Biermann battery could be obtained in torsioned cosmology. Nevertheless in this paper we show that this can be a particular result, since the second author did not took into account mean field dynamo equations in torsion field background. Actually it is shown that amplification or not of the magnetic field depends upon handness sign of the torsion field vector. It is shown that density fluctuations of spin-torsion density implies also a possibility of amplification of the cosmic magnetic fields. From WMAP data it is possible to estimate the spin-torsion fluctuation as $10^{-6}$ which represents an order of magnitude lower than the matter density. By making use of the gravitational couling of type $RF^{2}$ one obtains $10^{37}G$ for the Planck era magnetic field, which is a much more stringent limit than the ones obtained earlier. The magnetic field obtained today is $10^{-23}G$ is obtained which is able to seed galactic dynamos.
\end{abstract}
Key-words: Cosmological magnetic fields; mean field dynamos; torsion theories of gravity.
\newpage

\section{Introduction}
 Earlier Durrer et al \cite{1} have shown that is possible to obtain magnetic that may decrease during inflation but that can be amplified by dynamo mechanism. In this paper we show that in the realm of Einstein-Cartan inflationary cosmology the equipartition between magnetic and matter energy density implies that the spin-torsion density amounts an increase in the magnetic energy as a dynamo mechanism \cite{2}. Physically this unfortunatly is only obtained inside a high energy star such a black hole or in the very early universe. This result is obtained by the perturbation method of first order in density perturbations. Second-order in inflation scale is neglected. Here we show that this result may be obtained from Einstein-Cartan cosmology contrary to earlier results by Bamba et al \cite{3} and the author \cite{4}. This study which is undertaken in section 2 of the paper can be followed by a simple analysis of the mean field dynamo equations in torsioned background which takes into account the torsion fluctuations. Torsion is a very weak field and this result can certainly be better appreciated in high energy physics as is obtained in the early universe. Cosmological magnetic fields are shown to be amplified when torsion vector handness is negative while it can decay when handness of the torsion field is positive. Magnetic field fluctuations is also shown to be obtained from the interaction of the torsion field divergences with the magnetic field fluctuation itself. In the last section we consider the comoving cosmic flow and Minkowski spacetime ${\textbf{M}}^{4}$ plus torsion . However in previous section Hubble expansion and de Sitter inflation are consider. In the last section early universe dynamos \cite{1} are used to obtain a magnetic field for the Planck era as $B_{Planck}\sim{10^{37}G}$ and the cosmic magnetic field as today would be $B_{today}\sim{10^{-23}G}$ which is high enough to seed the galactic dynamos. Both limits are more stringent limits for the magnetic fields in the universe.
\section{Cosmological magnetic fields from Einstein-Cartan inflation}
In this section we shall shown that starting from Einstein-Cartan cosmological equations an early universe dynamo is possible in analogy with earlier results in GR by Durrer et al \cite{1}. The Einstein-Cartan-Friedmann equation is
\begin{equation}
H^{2}= \frac{8\pi}{3}\rho+{B^{2}}-\frac{k}{R}-\frac{2}{3}{\pi}^{2}{\sigma}^{2} \label{1}
\end{equation}
where $H=\frac{\dot{R}}{R}$ is the Hubble expansion factor and R is the universe radius. ${\sigma}^{2}$ here represents the spin-torsion energy density and k is the cosmological spatial curvature of the model. Here we shall be considering flat sections universe where $k=0$. Also B is the homogeneous magnetic field. By taking into consideration the fluctuation in the Hubble expansion as
\begin{equation}
H= H_{0}+{\delta}H
\label{2}
\end{equation}
and substitution into equation along with similar magnetic field fluctuations, one obtains
\begin{equation}
{H_{0}}^{2}=\frac{8\pi}{3}{\rho}[1+{\epsilon}_{B}-\frac{2\pi}{3}{\epsilon}_{\sigma}]
\label{3}
\end{equation}
Note that in the simple example of Einstein static universe, where $H_{0}$ vanishes, this relation reduces to
\begin{equation}
{\epsilon}_{B}=\frac{2\pi}{3}{\epsilon}_{\sigma}-1
\label{4}
\end{equation}
which shows clearly that the magnetic energy is enhanced by the spin-torsion density and an early universe dynamo is possible from strong torsion fields. This result is also valid when de Sitter inflation is obtained as a perturbation of the Einstein static universe with torsion and magnetic field. From expression (\ref{4}) we note that the magnetic energy density has to be weaker than spin-density which seems to be unfortunatly a very rare situation. To remedy this situation one can try to consider that $H_{0}={\alpha}$ corresponding to de Sitter metric
\begin{equation}
ds^{2}=e^{{\alpha}t}[dt^{2}-[dx^{2}+dy^{2}+dz^{2}]]
\label{5}
\end{equation}
Then the new Friedmann torsion equation would be
\begin{equation}
{\epsilon}_{B}={\epsilon}_{inflation}+\frac{2\pi}{3}{\epsilon}_{\sigma}-1
\label{6}
\end{equation}
where ${\epsilon}_{inflation}=\frac{{\alpha}^{2}}{\rho}$. Taking into account the fluctuation in H equation one obtains
\begin{equation}
\frac{{\delta}{H}}{\rho}=\frac{4\pi}{3}[\frac{{\delta}{\rho}}{\rho}+
\frac{{\delta}{{\rho}_{B}}}{\rho}-\frac{2\pi}{3}\frac{{\delta}{\rho}_{\sigma}}{\rho}]
\label{7}
\end{equation}

if one considers that the ratio on the LHS of this equation is neglected as in the early universe where the matter density ${\rho}$ is much bigger than the than the expansion of the universe in its infancy, this equation reduces to
\begin{equation}
\frac{{\delta}{\rho}}{\rho}=\frac{2\pi}{3}\frac{{\delta}{\rho}_{\sigma}}{\rho}-
\frac{{\delta}{{\rho}_{B}}}{\rho}
\label{8}
\end{equation}
From COBE WMAP data $\frac{{\delta}\rho}{\rho}\sim{10^{-5}}$. Taking into account that the magnetic field contrast $\frac{{\delta}{\rho}_{B}}{\rho}\sim{10^{-5}}$ \cite{4}, then the spin-torsion density can be easily computed as
\begin{equation}
\frac{{\delta}{\rho}_{\sigma}}{\rho}\sim{\frac{3}{\pi}\times{10^{-5}}}\sim{10^{-6}}
\label{9}
\end{equation}
Then the spin-torsion density contrast or fluctuation is one order of magnitude lower than the matter density fluctuation. On the contrary spin-torsion density fluctuation may affect the amplification of the magnetic field. In the next section we provide a more pedestrian way to understand the influence of torsion and its role in early universe dynamos.

\section{Cosmological magnetic fields from torsion handness}
 In this section we shall consider the comoving observer of the flow which allows us to work out with the Minkowski spacetime with torsion since torsion cannot be eliminated by equivalence Einstein principle such as the curved spacetime metric. Recently the author \cite{4} has shown that analogously to dynamo MHD equation in general relativity \cite{5}, is possible to obtain the dynamo equation in MHD relativistic background with torsion. This equation is given by
\begin{equation}
{\partial}_{t}\textbf{B}={\nabla}\times(\textbf{V}\times{\textbf{B}})+
\eta({{\nabla}.\textbf{T}})\textbf{B}+\eta{\nabla}^{2}\textbf{B}
\label{10}
\end{equation}
where $\textbf{T}$ and $\textbf{V}$ are respectively the torsion vector and the velocity of cosmic flow. Let us now to perform the fluctuation of these physical quantities as
\begin{equation}
\textbf{B}={\textbf{B}}_{0}+\textbf{b}
\label{11}
\end{equation}
where $\textbf{b}={\delta}\textbf{B}$, and
\begin{equation}
\textbf{V}={\textbf{V}}_{0}+\textbf{v}
\label{12}
\end{equation}
Actually \cite{2} the fluctuation in the magnetic field $\textbf{b}$ is given by
\begin{equation}
\textbf{b}(\textbf{x},t)=\sqrt{<{\textbf{b}}^{2}>}
\label{13}
\end{equation}
substitution of those expressions into the dynamo equations one obtains
\begin{equation}
\frac{{\partial}\textbf{b}}{{\partial}t}={\nabla}{\times}({\textbf{V}}\times{\textbf{B}}_{0})+
{\eta}[({\nabla}.\textbf{t}){\textbf{B}}_{0}+{\nabla}^{2}\textbf{b}]
\label{14}
\end{equation}
Here $\textbf{t}={\delta}{\textbf{T}}$ is the torsion fluctuation. By considering small turbulence the first term on the LHS of (\ref{14}) can be dropped, and this equation reduces to
\begin{equation}
{\nabla}{\times}({\textbf{V}}\times{\textbf{B}}_{0})+
{\eta}[({\nabla}.\textbf{t}){\textbf{B}}_{0}+{\nabla}^{2}\textbf{b}]=0
\label{15}
\end{equation}
This expression is already enough to demonstrate that the magnetic field fluctuations are generated by the interaction between the cosmic flow and the local mean field, and by interaction of the divergence of torsion with the fluctuation of the magnetic field itself. Now to further investigate the torsion action on cosmic magnetic fields let us consider the ansatz for the magnetic field fluctuation
\begin{equation}
{\textbf{b}}=Re{\textbf{b}}_{0}exp[i(\textbf{k}.\textbf{x}+{\omega}t)]
\label{16}
\end{equation}
Now let us perform the variation of the axial torsion ${\delta}T^{\mu}$ to obtain the last field equation for torsion
\begin{equation}
|\textbf{b}|=\frac{({\textbf{B}}_{0}.\textbf{k}){V}_{0}}{[{\omega}^{2}+{\eta}^{2}
(k^{2}+\textbf{k}.\textbf{t})^{2}]^{\frac{1}{2}}}
\label{17}
\end{equation}
Note from the denominator of this expression that the sign of handness of torsion fluctuation, given by the term $\textbf{k}.\textbf{t}$ determines whether the fluctuation amplifies the magnetic field or decays it. Of course when the handness is negative the magnetic field amplification is granted.
\section{Cosmological magnetic fields from coupling to curvature and torsion to seed galactic dynamos}
In this section we shall be concerned with the coupling between curvature and torsion straight from the electromagnetic field equations and possible determinations of the cosmological magnetic fields, taking also into account Riemann metric effects and not only torsion effects as in last section. We shall estimate a new and more stringent limit for the magnetic field at the Planck era. Here we adopt the Lagrangean
\begin{equation}
{\cal{L}}\sim{\sqrt{-g}[\frac{R}{2{\kappa}^{2}}+\frac{1}{4}F^{2}+ RF^{2}]}
\label{18}
\end{equation}
where $F^{2}=F_{{\mu}{\nu}}F^{{\mu}{\nu}}$ and R is the Ricci-Cartan scalar given by
\begin{equation}
R=R_{0}-\frac{1}{24}a^{2}S^{2}
\label{19}
\end{equation}
Here ${\kappa}^{2}$ is the Einstein gravitational constant, and $R_{0}$ is the Riemannian Ricci scalar given by
\begin{equation}
R_{0}=3\frac{\dot{H}}{a^{2}}+6\frac{H^{2}}{a^{2}}
\label{20}
\end{equation}
where $a^{2}$ is scale quantity which comes from the metric
\begin{equation}
ds^{2}=a^{2}(dt^{2}-[dx^{2}+dy^{2}+dz^{2}])
\label{21}
\end{equation}
Variation of the above Lagrangean with respect to the electromagnetic four-potential $A_{\mu}$ where $(0,1,2,3)$ one obtains the following equations
\begin{equation}
{\partial}_{\mu}[(1-4\frac{R}{{\kappa}^{2}})F^{{\mu}{\nu}}]=0
\label{22}
\end{equation}
The generalised Maxwell equations obtained from expressions (\ref{22}) in terms of the electric and magnetic vectors are given by
\begin{equation}
{\nabla}.\textbf{E}+{\nabla}ln{\psi}.\textbf{E}=0
\label{23}
\end{equation}
\begin{equation}
{\partial}_{t}\textbf{E}+{\partial}_{t}ln{\psi}\textbf{E}+{\nabla}{\times}\textbf{B}+{\nabla}ln{\psi}{\times}\textbf{B}=0
\label{24}
\end{equation}
\begin{equation}
{\nabla}.\textbf{B}=0
\label{25}
\end{equation}
\begin{equation}
{\partial}_{t}\textbf{B}=-{\nabla}{\times}\textbf{E}
\label{26}
\end{equation}
where
\begin{equation}
{\psi}:=[1-(4\frac{R}{{\kappa}^{2}}+\frac{1}{24}a^{2}S^{2})]
\label{27}
\end{equation}
The well-know electrodynamical expresion
\begin{equation}
\textbf{E}=-\textbf{V}{\times}\textbf{B}
\label{28}
\end{equation}
can be used to reduce the last Maxwell equation into the dynamo equation
\begin{equation}
{\partial}_{t}\textbf{B}={\nabla}{\times}[\textbf{V}\times\textbf{B}]
\label{29}
\end{equation}
Thus since only ${\psi}$ function depends upon torsion, one notices that this time dynamo equation does not depend on torsion, nevertheless is possible to show that indirectly torsion influences the magnetic field and causes its amplification. By considering that ${\psi}$ function is homogeneous, one simplifies considerably the above generalised Maxwell equations. Neverteheless, follwing Tsagas and Barrow´s idea \cite{5} that the primordial magnetic fields that can seed magnetic fields may be obtained from Maxwell traditional electrodynamics we shall consider that ${\psi}$ is constant, which from above equations reduce the couplings to classical Maxwell equations with the following constraint
\begin{equation}
({R_{0}}+\frac{1}{24}a^{2}S^{2})=constant
\label{30}
\end{equation}
which reduces to
\begin{equation}
[3\dot{H}+6H^{2}]=[c_{2}-\frac{1}{6}{S_{0}}^{2}]a^{2}
\label{31}
\end{equation}
which reduces to
\begin{equation}
\frac{\ddot{a}}{a}+\frac{{\dot{a}}^{2}}{a}=[c_{2}-\frac{1}{6}{S_{0}}^{2}]
\label{32}
\end{equation}
By taking into account the flux magnetic conservation as $B\sim{a^{-2}}$ which together $B\sim{t^{\alpha}}$ as usual for slow dynamos, and the equation
\begin{equation}
\frac{\ddot{a}}{a}+\frac{{\dot{a}}^{2}}{a^{2}}=\frac{1}{6}{S_{0}}^{2}
\label{33}
\end{equation}
yields for the cosmological magnetic field the following expression
Recalling that these equations are complex one obtains the following solution
\begin{equation}
B(t)\sim{{S_{0}}^{2}t}
\label{34}
\end{equation}
At the Planck era where $t_{P}\sim{10^{-43}s}$ the magnetic field is given by $B_{P}\sim{10^{37}G}$ which is a much more stringent value than the one obtained by Sivaram \cite{6} in Einstein-Cartan cosmology using f-meson dominance. Taking into account that the age of the universe is $t_{today}\sim{10^{17}s}$ the $B_{today}\sim{10^{-23}G}$ which can be useful to seed galactic dynamos. In the first case we have used a value of $S_{0}\sim{10^{-20}s^{-1}}$ and in the present universe $S_{0}\sim{10^{-17}cm^{-1}}$. Note that the very small value obtained by Sivaram of $10^{-38}G$ for the present universe was not due to the fact that he obtained this value from a much stronger estimate of $10^{58}G$ for the Planck magnetic field after the inflation but that he obtained this result from Einstein-Cartan cosmology which is utterly distinct than the $RF^{2}$ theory we use here.

\section{Discussions and Conclusions}
Earlier Sivaram  \cite{6} has investigated the cosmological magnetic fields in Einstein-Cartan cosmology making use of Einstein-Cartan cosmology and a relation between cosmic magnetic fields spin-density tensor and the rotation of the universe. This expression for the spin-torsion density yields a huge value for the magnetic field at Planck era which in turn decays after inflation yielding an extremely low value for the actual universe actually, as low as $10^{-38}G$ which cannot be used to seed any galactic dynamo. In the present paper using a distinct theory of cosmological torsion, based on $RF^{2}$ interaction Lagrangean, one obtains more stringent limits which seems able to seed the galactic dynamos after inflation. Besides we showed consistently that the sign of torsion handness is fundamental to decide whether and when torsion influences the amplification of cosmological magnetic fields. Actually torsion handness is a fundamental agent for the Lorentz violation \cite{7}. In this paper we also use the direct decoupling between torsion and electromagnetic fields, however a different choice can be made from the Maxwell generalised equations we derived in Section 3. This is a work in progress.
\section{Acknowledgements}
We would like to express my gratitude to M Dvornikov and A Brandenburg for helpful discussions on the subject of this paper. Financial support from CNPq. and University of State of Rio de Janeiro (UERJ) are grateful acknowledged.

\end{document}